\documentclass[12pt]{article}
\pdfoutput=1
\usepackage{graphicx}

\def\[{\begin{eqnarray}}
\def\]{\end{eqnarray}}
\def\d{\mathrm{d}}
\def\de{\delta}
\def\De{\Delta}
\def\ep{\epsilon}
\def\Ga{\Gamma}
\def\la{\lambda}
\def\om{\omega}
\def\th{\theta}

\def\O{\mathcal{O}}
\def\nn{\nonumber}

\begin{document}

\thispagestyle{empty}
\begin{flushright}\footnotesize
\texttt{}\\
\vspace{10mm}
\end{flushright}

\renewcommand{\thefootnote}{\fnsymbol{footnote}}
\setcounter{footnote}{0}

\begin{center}
{\Large\textbf{\mathversion{bold}
De (Baby) Sitter Overlaps}
\par}

\vspace{1.5cm}

\textbf{Marcus K. Benna} \\ \vspace{8mm} 
\textit{
Simons Center for Geometry and Physics, \\
State University of New York, Stony Brook, NY 11794-3636, USA 
} \\

\par\vspace{14mm}

\textbf{Abstract} \vspace{5mm}

\begin{minipage}{14cm}
In this note we employ methods borrowed from spin glass theory to study the phase space structure of fields in an inflating universe.
In particular, we compute the overlap distribution of a suitably coarse-grained, massless scalar on a $1+1$ dimensional (hence baby) de Sitter background, and find that (after an appropriate shift and rescaling) it is given by a Gumbel distribution.  We also calculate the triple overlap distribution of this system, whose characteristic function turns out to be a product of two Gumbel factors.
\end{minipage}

\end{center}

\vspace{0.5cm}

\newpage
\setcounter{page}{1}
\renewcommand{\thefootnote}{\arabic{footnote}}
\setcounter{footnote}{0}

\hrule
\tableofcontents
\vspace{8mm}
\hrule
\vspace{4mm}

\newpage

\section{Introduction}

We live in an expanding universe. There is very good evidence that this expansion is in fact accelerating \cite{Riess:1998cb, Perlmutter:1998np}, and the central idea of the theory of inflation \cite{Guth:1980zm, Linde:1981mu, Albrecht:1982wi} is that in the distant past, for a period of time not long after the big bang, this behavior was a dominant feature of our universe. 

During such periods of accelerated expansion, different regions of the universe lose causal contact with each other. A distant star we see in the sky one night might not be visible anymore the following night, not because it had stopped shining, but simply because the light it emits will never reach us.

On the other hand, we are used to thinking of physics in terms of local quantities and fields, which have correlators that fall off rapidly with distance. Let us consider an inflating universe endowed with a scalar field living in it. If this field is massless, its expectation value may vary widely from place to place.\footnote{Needless to say, there are no such massless scalars with significant couplings to the standard model particles in our universe today, else we would have noticed their profound consequences for low-energy physics (though almost massless scalar fields do occur e.g.~in models of slow roll inflation). Thus we should perhaps not think of this field as matter, but rather as an abstract probe of the properties of de Sitter space itself.}
Clearly, the field locally wants to minimize its energy density by not allowing steep gradients, but even small gradients can lead to vastly different expectation values over large distances. Once a neighboring region disappears behind our cosmological horizon, and we can no longer communicate with it, what is to stop the field in this region from drifting towards an expectation value completely different from what we measure in our vicinity? 
Thus small quantum fluctuations may be amplified into a wildly varying field profile. 

One might argue that a local observer should not care about expectation values in regions beyond his horizon.
But now let's take a more global point of view. Imagine an observer that looks back at these disconnected regions after inflation has ended (when formerly out of contact regions are becoming accessible again) and samples the average expectation values of the quantum field in each of those regions.  What is the distribution of these coarse-grained expectation values (and conjugate momenta) that such an observer would measure? 

Questions of this nature have been studied for a long time in the context of cosmological perturbation theory and structure formation \cite{Bardeen:1980kt, Mukhanov:1990me}. There is a large body of work seeking to predict what spectrum of matter density fluctuations a post-inflation observer should observe in the sky, assuming they arise from primordial quantum fluctuations and are modified according to the details of one's favorite inflationary model.

But here we would like to take a further step back, and consider the statistical properties of these coarse-grained variables in phase space. 
Given the thermal nature of the de Sitter vacuum, is there ergodicity breaking? Does this system exhibit a glassy structure, as suggested in \cite{AD}?

We would like to shed some light on the global structure of the reduced phase space of expectation values (and conjugate momenta) corresponding to causally disconnected regions on a constant time slice. To this end, we will make use of tools and techniques from spin glass theory (see e.g.~\cite{MPV,CG,DG}), in particular the overlap and triple overlap distributions, which can be understood as order parameters for a non-trivial structure in the space of thermodynamic states \cite{Parisi:1983dx}.

The application of spin glass methods to de Sitter dynamics was pioneered by F.~Denef and D.~Anninos \cite{AD, Denef:2011ee, APhD, Unpub}, who conceived the central ideas and were the first to investigate overlaps of various types on de Sitter spaces of different dimensionalities. They developed the formalism of reduced density matrix based overlaps that we will employ, and also wrote down the simple model that we will study, as an ideal playground to both test these techniques and attempt to tackle questions about the state space of de Sitter. In Section \ref{RevSet} we will review the relevant parts of their work, and remind the reader of some simple definitions and statistical notions borrowed from the theory of spin glasses. 

Section \ref{Low} describes a direct attempt to characterize the overlap, which will give us the first few moments of the distribution, while in Section \ref{Graphs} we introduce a slightly more sophisticated method, based on graphs and combinatorics, that allows us to find the complete distribution (at leading order in a late-time expansion). In Section \ref{Triple} we generalize this method to compute the triple overlap distribution, before concluding with a brief discussion of our results.

\section{Setup and Review} \label{RevSet}

\subsection{A Simple Model}

The model we will consider \cite{Unpub} could hardly be any simpler. As a background will take a 1+1 dimensional de Sitter spacetime, whose metric we can write in global coordinates as
\[
\d s^2 = - \d \tau^2 + \cosh^2 \tau\, \d \th^2 \ ,
\]  
where we have set the cosmological constant to unity.
A fixed time $\tau=$ const.~slice is then simply a circle parametrized by the angular variable $\th \in (0,2\pi)$.  
On this background we study the evolution of a real, massless scalar field $\phi(\tau,\th)$. For a constant time slice we can Fourier decompose its fluctuations in the obvious fashion\footnote{We have chosen a closed slicing here in order to obtain discrete modes.}
\[
\phi = \sum_{n=-\infty}^\infty \phi_n\, e^{i n \th} \ ,
\]
where reality requires that $\phi_{-n} = \phi_n^\ast$. The Euclidean vacuum wavefunctional \cite{HartHawk, Allen:1985ux} can then be written as 
\[\label{EucVac}
\Psi = \prod_{n=1}^\infty \sqrt{2 n \over \pi}\, \exp\left(- n |\phi_n|^2 \right) \ ,
\] 
which simply takes the form of an infinite sequence of harmonic oscillators (and we have removed the zero mode).\footnote{This is closely related to the string world-sheet vacuum studied in \cite{Karliner:1988hd}.}

All this is very elementary and its seems hard to believe that anything interesting could come out of it. 
However, as we have remarked above, a local observer only has access to his own causal diamond, and thus cannot measure all of the Fourier modes of the scalar field. 

The Penrose diagram of de Sitter is of course simply a rectangle, and the closer we are to future space-like infinity (the top of the diagram) the more causally disconnected regions there are on a constant $\tau$ (which implies also constant conformal time) slice. Let's consider such a slice with $\mathcal{O}(1/\ep)$ regions of typical size $\ep$. Our future observer that looks back long after inflation has ended can in principle measure the average state (expectation value and momentum) inside all of these boxes\footnote{In the case of a strict de Sitter universe inflation of course never ends, but we could imagine an unphysical, omniscient being that has access to the expectation values of the field inside all the causally disconnected regions on a space-like hypersurface. This latter point of view is of course not just unphysical in that no such observer exists, but it also opens up a Pandora's box of dicey questions concerning the interpretation of quantum mechanics in the context of cosmology. We will not discuss such issues here, so our discussion of de Sitter should be understood merely as an idealization approximating the post-inflation observer's point of view.}, and may choose to introduce corresponding coarse grained variables defined by convoluting the field profile with a suitable window function.  

Let us choose for this purpose a Lorentzian window (or box) function centered at $\th_i$
\[ \label{Box}
\mathcal{B}_i(\th - \th_i) = \sqrt{\ep^3 \over 4 \pi}\, {1 \over (\th - \th_i)^2 + (\ep/2)^2} \ ,
\]
since its Fourier transform is conveniently given by an exponential, and thus the average value of the field inside this box is given by the convolution
\[ 
\int_{-\infty}^{\infty} \d\th\, \mathcal{B}_i(\th - \th_i)\, \phi(\th) = \sqrt{ \pi\ep} \sum_{n=-\infty}^{\infty}  e^{i n \th_i} e^{-n \ep/2}\, \phi_n \ .
\]
Here we have taken the range of integration as infinite, which is an approximation that becomes exact in the limit of vanishing $\ep$ which we will be interested in below (or alternatively we can think of the integral as simply wrapping around the circle).

Note that while the vacuum (\ref{EucVac}) is a pure state in the full Hilbert space, this is no longer true in the reduced state space resulting from coarse-graining.
What we are going to be concerned with are the statistical properties of these coarse-grained variables, and it is their overlap distribution we would like to compute. First, however, we will have to define this distribution and introduce some techniques for calculating it. 

\subsection{Review of Overlap Distributions}

Let us briefly recall the basic concept of an overlap distribution and some useful definitions, as explained in \cite{MPV,CG,DG, Parisi:1983dx} and the excellent recent lecture notes \cite{Denef:2011ee}.  

In complex materials a thermodynamic state described by a density matrix $\rho$ may at low temperatures break ergodicity (i.e.~not all accessible microstates or phase space cells will be equally likely). This can occur as a trivial consequence of symmetry breaking (as in a ferromagnet), but in more non-trivial systems such as spin glasses ergodicity breaking takes place irrespective of any obvious associated symmetry.  We should then be able to decompose the density matrix into different ergodic components (sometimes referred to as thermodynamic pure states)
$\rho = \sum_\alpha w_\alpha \rho_\alpha$. 

Different ergodic components in phase space are separated by very high free energy barriers, and thus no longer communicate with each other at low temperatures.  
The overlap $q_{\alpha\beta}$ between two such components is then defined by averaging over all degrees of freedom the product of their expectation values in states $\alpha$ and $\beta$. 
E.g.~for a system of $N$ spins $s_i$, we would write $q_{\alpha\beta} \equiv {1 \over N} \sum_i \langle s_i \rangle_\alpha \langle s_i \rangle_\beta$.

Usually it is not possible to explicitly construct the decomposition into thermodynamic pure states, but a quantity of interest that may be computable is the overlap distribution
\[ \label{OverlapDef}
\mathcal{P}(q) \equiv \sum_{\alpha,\beta} w_\alpha w_\beta\,  \de(q - q_{\alpha \beta}) = \left\langle \de\left( q - {1 \over N} \sum_i s_i^1  s_i^2 \right) \right\rangle_{2\ \mathrm{replicas}} \ .
\]  
The last equality is a rather non-trivial step, that forms the centerpiece of the replica formalism. It re-expresses  the sum over ergodic components (which we cannot perform explicitly) as an expectation value of an overlap between two copies of the system. We will not reproduce the proof of this statement here, but instead refer the reader to the literature for justification \cite{MPV, Denef:2011ee}.\footnote{Note that there is nothing unphysical about working with two copies of the same system, and that this should not be confused with the so-called replica trick for computing quenched averages of the free energy, which employs an arbitrary number of replicas and then analytically continues that number to zero.}

As we keep lowering the temperature, the breaking up of the phase space may continue in several steps, leading naturally to a tree structure in the space of states. This is best probed by computing the triple overlap distribution
\[
\mathcal{P}(q_1,q_2,q_3) \equiv \sum_{\alpha,\beta,\gamma} w_\alpha w_\beta w_\gamma \, \de(q_1 - q_{\beta\gamma})\, \de(q_2 - q_{\gamma \alpha}) \, \de(q_3 - q_{\alpha \beta}) \ ,
\]
which can be similarly expressed in terms of a three replica expectation value. If a hierarchical tree structure is present in the space of states it will be reflected in this quantity exhibiting ultrametricity \cite{MV}, namely the property that out of any triple of overlaps $(q_1,q_2,q_3)$ the largest two will be equal, as is the case for pairwise path-distances between any three leaves of a tree. This means that the distribution will have support only on a codimension one subspace where the three $q_a$ form equilateral or isosceles triangles.

\subsection{Introducing Wigner Densities}

Now let's consider more generally $N$ degrees of freedom living in a Hilbert space $\mathcal{H}$, and denote by $\mathcal{H}_i$ the Hilbert space associated with one of them. We can find a reduced density matrix $\rho^i$ by tracing over all other degrees of freedom, and use this to define a new kind of overlap\footnote{We use the upper case symbol $Q_{\alpha\beta}$ here to stress the analogy with $q_{\alpha\beta}$ in the spin example, but note that the range in which this overlap takes values is shifted and rescaled compared to the definition above.}, first introduced in \cite{Denef:2011ee},
between two states labelled by $\alpha$ and $\beta$ as
$Q_{\alpha\beta}  \equiv {1 \over N} \sum_i \mathrm{Tr}_{\mathcal{H}_i}\, \rho^i_\alpha \rho^i_\beta$.

We can recast this overlap in a very suggestive form by working with Wigner distributions instead of density matrices. Recall that the Wigner distribution in phase space is defined as a Fourier transform of certain matrix elements of the density operator. In the case at hand, for a scalar field $\phi$ with momentum $p$, in the Euclidean vacuum state $\rho = |\Psi\rangle \langle \Psi |$, we find
\[ 
W(\phi, p) \equiv \int \d s\, e^{i s.p}\,  \langle \phi + s/2 | \rho | \phi - s/2 \rangle = \prod_{n=1}^\infty 4\, e^{-2n |\phi_n|^2 - |p_n|^2 /(2 n)} \ , 
\]
where $s$ and $p$ satisfy the same reality condition as $\phi$, and the integral measure is $\d s = \prod_{n=1}^\infty \d s_n$, with the $s_n$ integrated over the whole complex plane. Here we have set Planck's constant to unity and used the shorthand $s.p = {1 \over 2}\sum_{n = -\infty}^\infty s_n\,  p_n$, again with the zero-mode removed. 

We can also write down the Wigner densities $W^i$ corresponding to the reduced density matrices for the average expectation value $X_i$ and conjugate momentum $P_i$ of the field inside a certain region $\mathcal{B}_i$ \cite{Unpub}
\[ \label{RedWig}
W^i = \int \d \phi\, \d p \,\de\left(X_i - \int_{-\infty}^{\infty} \d\th\, \mathcal{B}_i(\th - \th_i)\, \phi(\th)\right)\, \de\left(P_i - \int_{-\infty}^{\infty} \d\th\, \mathcal{B}_i(\th - \th_i)\, p(\th)\right) \, W(\phi, p)  \ , \quad
\]
where the integration measure is $(2 \pi)^{-2} \prod_{n=1}^\infty \d \phi_n \, \d p_n$ and the range again extends over the whole complex plane.\footnote{The coarse-graining involved in the definition of the reduced Wigner densities reflects the thermal nature of the cosmological horizon of a localized observer; in particular, the associated density matrices no longer correspond to pure quantum states such as the Euclidean vacuum state (\ref{EucVac}) we stated with.}

Using these reduced Wigner distributions on the phase space for one particular degree of freedom, we can recast the overlap in the following form \cite{Denef:2011ee}
\[
Q_{\alpha\beta} = {1 \over N} \sum_i \int \d X_i \, \d P_i \, W^i_\alpha(X_i, P_i) \, W^i_\beta(X_i, P_i) \ ,
\]
except that in the calculations below we will not actually work with $N$ fixed boxes in positions $\theta_i$ and average over them, but instead replace this sum by an integral $(2 \pi)^{-1} \int_0^{2\pi} \d \th_i$.

This expression for the overlap can now be used to rewrite the overlap distribution as a two replica expectation value, in complete analogy with the right hand side of (\ref{OverlapDef}). While an expectation value of a delta function may appear to be a somewhat singular representation, the important implication is that the moments of the distribution \cite{Unpub} are given by
\[ \label{MomDef}
\langle q^K \rangle = \left(\prod_{i=1}^K \int_0^{2 \pi} {d \th_i \over 2 \pi} \int_{-\infty}^\infty \d X_i \, \d P_i  \right) \left(W^{123 \ldots K}\right)^2 \ ,
\]
where $W^{123 \ldots K}$ is a multivariable Wigner density defined exactly as in (\ref{RedWig}) except with $K$ pairs of delta function insertions  
that set the average expectation values of the field $\phi(\th)$ inside the boxes at positions $\th_i$ equal to $X_i$ and similarly equate the average conjugate momenta to $P_i$.

For our particular choice of window function (\ref{Box}) we can find the reduced Wigner distribution by using the integral representation of the $K$ pairs of $\de$-functions with integration variables $\la_i$ for expectation values and $\tilde{\la}_i$ for momenta, which gives 
\[ \label{Wigner}
 W^{123\ldots K} &=& \int_{-\infty}^{\infty} \left(\prod_{k=1}^K \, \d\la_k\,  \d\tilde{\la}_k \, \ e^{2\pi i (\la_k X_k + \tilde{\la}_k P_k)} \right) \\
&&\times \exp \left( - 4 \pi^3 \ep \sum_{n=1}^{\infty} e^{- n \ep} \sum_{i,j = 1}^{K} \left({\la_i \la_j \over 2 n} + 2 n \tilde{\la}_i \tilde{\la}_j \right)  \cos n (\th_i - \th_j)  \right) \ . \nn
\]

Provided we can compute from (\ref{MomDef}) the moments for all integers $K$, we can find the characteristic function, and hence reconstruct the full overlap distribution $\mathcal{P}(q)$.
If the system has a non-trivial phase structure, which is related to failure of cluster decomposition as discussed in \cite{Denef:2011ee}, we should find a non-trivial overlap distribution. In this sense this quantity can be understood as an order parameter for ergodicity breaking.

\section{A First Stab at the Overlap Distribution} \label{Low}

Having explained the basic setup of the computation, let us now tackle the calculation of the first few moments of the overlap distribution.
From the analogy with complex materials in the previous section, it should be apparent that the box size $\ep$ plays a role analogous to temperature, and therefore we will be interested in the small $\ep$ (i.e.~late conformal time) expansion in which there are many causally disconnected regions in our model universe.

\subsection{First Moment}

In order to find the first moment of the overlap distribution, we merely have to simplify the expression (\ref{Wigner}) for the reduced Wigner density with one index. There are no angular integrals to be performed (since by rotational symmetry the moments only depend on differences of angles), and the integrals over $X_1$ and $P_1$ simply lead to delta functions that set to zero the sums of the $\la_i$ and $\tilde{\la}_i$ parameters, respectively, for the two copies of the Wigner density.
Evaluating the sums $\sum_{n=1}^{\infty} 2 n \, e^{- n \ep}  = (\cosh \ep -1)^{-1}$ and $\sum_{n=1}^{\infty} e^{- n \ep}/(2n)  = -{1\over 2} \log(1-e^{-\ep})$ we find that the first moment is given by
\[
\langle q \rangle &=& \int \d X_1 \d P_1 \, (W^1)^2 = \int_{-\infty}^{\infty}  \, \d\la_1 \d\tilde{\la}_1 \exp \left( - 8 \pi^3 \ep  \left( -{\la_1^2 \over 2} \log(1-e^{-\ep}) + { \tilde{\la}_1^2 \over \cosh \ep -1 } \right)  \right) \nn \\
&=& { \sinh {\ep \over 2} \over 4 \pi^2 \ep \sqrt{ -\log(1-e^{-\ep})}} \ \stackrel{\ep \to 0}{ \longrightarrow } \ {1 \over 8 \pi^2 \sqrt{-\log \ep}} + \mathcal{O} (\ep(-\log\ep)^{-3/2}) \ .
\]

Note that the leading term is proportional to $(-\log\ep)^{-1/2}$, which tells us that the average overlap goes to zero (albeit rather slowly) with conformal time. The precise coefficient of $\langle q \rangle$ depends our choice of window function (i.e.~on how many boxes of a given shape and width $\ep$ can effectively fit into our one-dimensional universe), and thus is not particularly physical, though it will be crucial when comparing to higher moments of the distribution.

\subsection{Second Moment}

To compute higher moments we have to perform angular integrals over the (differences in) box positions.
For $K=2$ there is only one independent angle (since one can always be fixed using the rotational symmetry) and the integral can be evaluated exactly.

Using the sums
\[ 
\label{SumP}
\sum_{n=1}^{\infty} 2 n \, e^{- n \ep} \cos(n \Delta\th)   &=& {\cosh \ep \cos \Delta\th -1 \over \cosh \ep - \cos \Delta\th} \ ,  \\
\label{SumX}
\sum_{n=1}^{\infty} {e^{- n \ep} \over 2n} \cos(n \Delta\th)  &=& -{1\over 4} \log(1-2 e^{-\ep} \cos \Delta\th + e^{-2\ep}) \ ,
\]
we find that the second moment is given by
\[
\int \d X_1 \d P_1 \d X_2 \d P_2 \, (W^{12})^2 = {1 \over 64 \pi^4 \ep^2} { 1 \over \sqrt{\det M}}{ 1 \over \sqrt{\det \tilde{M}}} \ ,
\]
where the symmetric matrix $\tilde{M}$ has entries $\tilde{M}_{ij}$ given by equation (\ref{SumP}) with $\Delta\th = \th_i - \th_j$, and similarly the symmetric matrix $M$ has matrix elements given by equation (\ref{SumX}).\footnote{We should remark at this point that in principle any reasonable (i.e.~continuous and square integrable) window function could have done do the job instead of (\ref{Box}), and suitably regularized the computation of the matrix elements of $M$ and $\tilde{M}$. However, the Lorentzian is particularly convenient, since it allows us to explicitly perform the resulting sums (\ref{SumP}) and (\ref{SumX}) which makes the following computations much more manageable. We have tried out different box functions, and found that the crucial long-range behavior of (\ref{SumX}) was identical, but accompanied by various short range terms modifying this for small $\Delta\th$. If we grant that the long range behavior is universal, the detailed choice of box function should not matter for subsequent calculations of the shape of the (appropriately rescaled) distribution in the small $\ep$ limit.}

Due to the smoothing provided by the window function, all matrix elements are finite, and divergences can only occur when one of the matrices develops a zero eigenvalue (such that the determinant vanishes) which happens when the two boxes coincide. This singularity is not integrable, thus we have to be careful to exclude from the integral a region in which the angular separation becomes less than $\mathcal{O}(\ep)$.\footnote{A different method to treat coincident boxes, which may be conceptually clearer, is as follows. Instead of defining the the reduced Wigner distribution with an insertion of $\de(X_i - \int \d \th \, \mathcal{B}_i \phi(\th))$, we can broaden the delta function to a Gaussian of width $\de$, i.e. instead of demanding that the averaged field be exactly equal to a specified value, we only ask for approximate (smeared) agreement. It is easy see that this ensures that matrix $M$ never degenerates, i.e. has non-vanishing determinant even for coincident boxes. The same is true for the matrix $\tilde{M}$ with the corresponding broadening of the momentum space delta function. We should then compute the moments of the overlap distribution as functions of $\ep$ and $\de$ and in the end take the double limit of both small parameters going to zero. Unfortunately this makes the calculation somewhat unwieldy in practice.}
 
It is not hard to see that the momentum determinant  can be expanded in a power series in $\ep$
\[
{ 1 \over \sqrt{\det \tilde{M}}} = {\ep^2 \over 2} + \mathcal{O}(\ep^4) \ .
\]
Furthermore, even though subleading terms in this expression diverge for vanishing angular separation $\Delta\th =0$, after performing the angular integral over the interval $(\ep, 2\pi - \ep)$, say, the contribution of the momentum determinant remains $\ep^2/2$ to leading order.   
Thus we have to focus on the position determinant, and we will see below that this behavior continues for the computation of higher moments: the momentum determinant contributes only a constant, since the momentum correlators are short-ranged.  

It is convenient to rescale the parameters $\la_i$ by $(-4\pi^3 \ep \log(1-e^{-\ep}))^{-1/2}$ so that the diagonal terms become $\ep$ independent Gaussians, and 
\[ \label{qsquared}
\langle q^2 \rangle = {1 \over 8 \pi^2 \ep} {\ep^2 \over 2} \int {\d \De\th \over 2 \pi} \int_{-\infty}^{\infty}  \, {\d\la_1 \d \la_2 \, e^{-\la_1^2 -\la_2^2 } \over - 4 \pi^3 \ep \log(1-e^{-\ep}) } \left( 1 - 2 e^{-\ep} \cos \De\th + e^{-2\ep}\right)^{\la_1 \la_2 \over -\log(1-e^{-\ep})} \ .
\]

We are now supposed to integrate over the parameters $\la_i$, and in the end over the angle $\De\th$, being careful to exclude a small region around $\De\th=0$. This is not very practical however, and we will instead switch the order of integration. 

Performing the angular integral first, over the full range $(0,2\pi)$, which in this case can be done explicitly, leads to a hypergeometric function depending on $\ep$ and the combination $c \equiv {-\la_1 \la_2 / \log(1-e^{-\ep})} $ appearing in the exponent above. Of course, if we then try to integrate over the parameters $\la_i$ we again encounter the singularity originating from coincident boxes. For very negative values of the exponent $c$ the hypergeometric function grows so fast that the Gaussian envelope in (\ref{qsquared}) fails to render the integral finite. 

It would appear that we have not gained anything, and still have to introduce a regulator to obtain a well-defined answer. However, at this point we recall that we are really interested in the thermodynamic limit, which in this case means the limit of small $\ep$. At late cosmological times, the (causally disconnected) boxes become very small, and two randomly chosen ones are very unlikely to be coincident.

Taking the small $\ep$ limit after carrying out the angular integral in (\ref{qsquared}) allows us to separate the divergent part of the answer from the $\O(\ep^0 (\log \ep)^n)$ terms that we are interested in.\footnote
{
More precisely, if we consider the resulting hypergeometric function for small $\ep$, but keeping $c$ constant, the expansion we find is not uniform. It is the sum of a power series starting at $\O(\ep^0)$ and another power series that contains a factor $\ep^{2c}$ and starts at $\O(\ep^{2c+1})$. The latter contains the singularity at large negative $c$, and we drop it, while the former describes the behavior at small $c$ (it dominates for $c > -1/2$), and thus we keep its leading term. Remembering that $c$ goes to $-\la_1\la_2/\log \ep$ it is clear that the physically interesting region is that of small $c$. 

Put differently, the terms that lead to divergences in the $\la_i$ integral are $\O(\ep^1)$ with a prefactor $e^{-2 \la_1 \la_2}$, while the terms we are interested are present at zeroth order in the $\ep$ expansion (with correction in powers of $\log \ep$), but don't contain the offending exponential factor. 
} 

Note that of course one cannot simply set $\ep$ to zero from the start. It is crucial to extract the dependence on $\log \ep$ first, which is most easily done by rescaling the $\la_i$ appropriately. In fact, going back to (\ref{qsquared}) one can check that performing the integral for finite $\ep$ and then taking the thermodynamic limit is equivalent to letting $\ep$ go to zero in the integrand, such that it appears only inside inside the logarithms, and then performing the integral.  

This leads to an expansion of the second moment in inverse powers of $\log\ep$
\[
\langle q^2 \rangle = { 1 \over -64 \pi^4 \log\ep} \left( 1+ {\pi^2 \over 24 \log^2 \ep} + {19 \pi^4 \over 640 \log^4 \ep} +{1375 \pi^6 + 151200 \zeta^2(3) \over 21504 \log^6 \ep} + \O(\log^{-8} \ep )\right) \ . \qquad
\]
Higher order terms are easily computed by expanding further. Note that this expansion satisfies a transcendentality principle; each inverse power of $\log\ep$ comes with a power of $\pi$ (or corresponding $\zeta$ value). Also, the leading term is precisely equal to $\langle q \rangle^2$.
  
In summary, our prescription for computing the overlap, which we will also apply to higher moments below, is to first rescale the auxiliary parameters $\la_i$ and take the thermodynamic limit, throwing away terms proportional to powers of $\ep$ (which may have led to divergences), but being careful to retain the dependence on $\log \ep$. After that, we perform the angular integral(s) over the full interval from $0$ to $2\pi$, and in the end integrate over the (rescaled) $\la_i$.    
  
\subsection{Third Moment}

Computing higher moments of the overlap distribution requires us to perform multiple angular integrals, while being careful to treat coincident boxes properly. This can again be formulated as finding the determinants of $M$ and $\tilde{M}$, which for the $K$th moment are $K$ by $K$ matrices, with the off-diagonal matrix elements  $M_{ij}$ and $\tilde{M}_{ij}$ depending on the angular separation $\th_i - \th_j$, again given by (\ref{SumX}) and (\ref{SumP}).

Hence for $K=3$ we have 
\[
\int \prod_{i=1}^3 \d X_i \d P_i \, (W^{123})^2 = {1 \over (8 \pi^2 \ep)^3} { 1 \over \sqrt{\det M}}{ 1 \over \sqrt{\det \tilde{M}}} \ .
\]

As before one can show that the momentum determinant contributes only a constant factor, in this case ${ 2^{-3/2}\ep^3}$. Rescaling the $\la_i$ and taking the $\ep \to 0$ limit then leads to the expression
\[ \label{qcubed}
\langle q^3 \rangle &=& {1 \over (4 \pi)^3}  \int_0^{2\pi} {\d \th_1 \d \th_2 \d \th_3 \over (2 \pi)^3} \int_{-\infty}^{\infty}  \, {\d\la_1 \d \la_2  \d \la_3 \, \over (- 4 \pi^3 \log \ep)^{3/2} } \ e^{-\la_1^2 -\la_2^2 -\la_3^2 }  \nn \\
&&\times \left(4\sin^2 { \theta_1-\theta_2 \over 2}\right)^{-{\la_1 \la_2 \over \log \ep}}  \left(4\sin^2 { \theta_1-\theta_3 \over 2}\right)^{-{\la_1 \la_3 \over \log \ep}} \left(4\sin^2 { \theta_2-\theta_3 \over 2}\right)^{-{\la_2 \la_3 \over \log \ep}} \ .
\]

While it is not obvious a priori how to compute this integral exactly for arbitrary exponents, we can try to use a procedure similar in spirit to the replica trick: find an appropriate function that agrees with this integral for integer exponents (in which case the integral can be performed explicitly) and then use this function to extrapolate to small values of the exponent. This is discussed in Appendix \ref{K3Int}.

Using the result (\ref{3AngInt}) from the appendix we find that the third moment is given by
\[
\langle q^3 \rangle =  {1 \over (-64 \pi^4 \log\ep)^{3/2}} &&\left( 1+ {\pi^2 \over 8 \log^2 \ep} + {\zeta(3) \over 4 \log^3 \ep} + {67 \pi^4 \over 640 \log^4 \ep} \right. \nn \\
&&\left. +{9 \pi^2 \zeta(3) + 108 \zeta(5) \over 32 \log^5 \ep}+ \O(\log^{-6} \ep )\right) \ .
\]

Again this obeys the transcendentality principle, and one can easily compute further terms if one so desires.

\subsection{The Lowest Central Moments}

We have already noted that to leading order $\langle q^2 \rangle = \langle q \rangle^2$ and thus the second central moment (variance) is actually of order $\log^{-3} \ep$:
\[
\langle q^2 \rangle_c =  {1 \over -64 \pi^4 \log\ep} \left( {\pi^2 \over 24 \log^2 \ep} + {19 \pi^4 \over 640 \log^4 \ep} + \O(\log^{-6} \ep ) \right ) \ .
\]  

More surprisingly, computing the third central moment we find that the first two leading terms cancel out
\[
\langle q^3 \rangle_c = \langle q^3 \rangle - 3\langle q^2 \rangle \langle q \rangle +2 \langle q \rangle^3 = {1 \over (-64 \pi^4 \log\ep)^{3/2}} \left( {\zeta(3) \over 4 \log^3 \ep} + {\pi^4 \over 64 \log^4 \ep} + \O(\log^{-5} \ep )\right)  \ . \qquad
\]

This raises the question of whether the $K$th central moments $\langle q^K \rangle_c$ are all of $\O(\log^{-3K/2} \ep)$.
If this is the case there might be a closed form expression for the leading overlap distribution (i.e. capturing the leading term of each central moment) as a function of $\hat{q} \sim (\log^{3/2} \ep) (q - \langle q \rangle) $, without explicit $\ep$ dependence.

In fact, the above cancellations suggests that perhaps one should set up the calculation directly for central moments, rather than computing ordinary moments and subtracting. Certainly, if we were to use numerical methods this would be absolutely necessary, and  
we will see below that this also leads to substantial simplifications in the exact analytic computation.

\subsection{Fourth and Higher Moments}

The computation of higher moments is formally very similar to what we have discussed above. 
The crucial integral that needs to be evaluated to find the $K$th moment of the overlap distribution is given by\footnote{Essentially this amounts to studying a one-dimensional model of particles on a circle with a pairwise interaction potential given by the logarithm of the chordal distance, and arbitrary real-valued charges $\la_i$.}
\[ \label{HigherInt}
\int_0^{2\pi} \prod_{k=0}^K {\d\th_k \over 2 \pi} \prod_{j>i \ge 1}^K \left(4 \sin^2 { \th_i-\th_j \over 2}\right)^{-\la_i \la_j / \log\ep} \ .
\]
Knowledge of this integral for all positive integers $K$, even if only as an expansion for small $\ep$, would allow us to reconstruct the complete overlap distribution.

However, already for $K=4$ this is rather non-trivial. Appendix \ref{K4Int} describes the treatment analogous to the one that worked for $K=3$,  while Appendix \ref{Modest} discusses what we get if we expand in $\ep$ first. It is clear from these discussions that it would be difficult to directly compute arbitrary higher moments in this fashion, and we instead have to think of a smarter method to achieve this.

\section{Central Moments and Graphs} \label{Graphs}

From the above we know that the overlap distribution is approximately a delta-function centered on $\langle q \rangle =  (-64 \pi^4 \log\ep)^{-1/2}$, and that we should really compute central moments to learn more about its shape. 

\subsection{Expansion in Terms of Complete Graphs}

The central moments are given by 
\[
\langle q^K \rangle_c &\equiv& \langle (q-\langle q \rangle)^K \rangle = \sum_{n=0}^K {K \choose n} \langle q^n \rangle (-\langle q \rangle)^{K-n}   \\
&=& \langle q^K \rangle - K \langle q^{K-1} \rangle \langle q \rangle + {1\over 2} K (K-1) \langle q^{K-2} \rangle \langle q \rangle^2  - \ldots + (-1)^{K-1} (K-1) \langle q \rangle^K  \ . \nn
\]

Every term in this expansion, when written out explicitly in terms of Wigner functionals as in the previous section, will boil down to a certain angular integral of the type discussed above, which we can represent symbolically by a graph. If draw a vertex labelled by $i$ for the angular integral $\int \d \theta_i / (2 \pi) $ and an (undirected) edge between vertices $i$ and $j$ for each factor of $(4 \sin^2 { \th_i-\th_j \over 2})^{-\la_i \la_j / \log\ep}$, then $\langle q^K \rangle$ corresponds to the complete graph with $K$ vertices (i.e. the graph in which each vertex is connected to every other one by exactly one undirected edge). We will denote this graph by $\mathcal{C}_K$.

The next term in the expansion above has a factor of $\langle q \rangle$, which corresponds to an integral over an angle that nothing depends on, which is represented by a vertex that has no edges attached to it. This gives simply a pure number, which after factoring out a suitable overall coefficient is unity. The factor $\langle q^{K-1} \rangle$ on the other hand again corresponds to a complete graph, but with only $K-1$ vertices. There are $K$ such terms, and we can think of these as arising from the $K$ complete graphs  $\mathcal{C}_{K-1}$ that are subgraphs of  $\mathcal{C}_K$.

Similarly, all subsequent terms in the above expansion can be associated to complete graphs of fewer vertices, and the combinatorics is such that there is precisely one term of coefficient one or minus one for every complete graph that is a subgraph of $\mathcal{C}_K$. Finally, the last term is just a number, $(-1)^{K-1} (K-1)$, indicating graphs with no edges, which we can think of as arising from  $K$ complete graphs with one vertex (singletons) and one graph with no vertices (the null graph). 

Why is this  relevant? Naively the $K$th moment will be of $\O(\log^{-K/2} \ep)$, but we have seen above that due to certain cancellations at the lowest central moments are actually parametrically smaller than that. The graphical representation can help us rewrite the computation in a way the makes these cancellations manifest for all $K$, and allows us to easily identify the leading pieces of the answer in the small $\ep$ limit.

\subsection{Expansion with Parametrically Small Edge Factors}\label{EdgeExp}

We will illustrate the idea with the simple example of $K=3$. If we introduce the shorthand $(ij) \equiv (4 \sin^2 { \th_i-\th_j \over 2})^{-\la_i \la_j / \log\ep}$ for the angular factor associated to the edge between vertices $i$ and $j$, and declare it as understood that all angles will be integrated over with measure $\int_0^{2 \pi} \d \theta_i / (2 \pi) $, and all parameters $\la_i$ with measure $\int_{-\infty}^{\infty} \d \la_i \exp(-\la_i^2) / (\sqrt{\pi}) $, we have
\[
(-64 \pi^4 \log\ep)^{3/2} \langle q^3 \rangle_c &=& (12)(13)(23)  - (12) - (13) - (23) + 2  \\ 
&=&  [(12)-1][(13)-1][(23)-1]  \nn \\
&+&  [(12)-1][(13)-1] + [(12)-1][(23)-1]  + [(13)-1][(23)-1]  \ . \nn
\]
Here we have rewritten the polynomial in $(ij)$ in terms of shifted variables, which we will denote by $[ij] \equiv (ij)-1$. While the  $(ij)$ are of $\O(1)$, the factors of $[ij]$ are of $\O(\log^{-1}\ep)$, which makes it obvious that the right hand side is actually of $\O(\log^{-3}\ep)$ and that the leading contribution comes entirely from the term cubic in the rectangular brackets. 

To see this, we merely have to consider the small $\ep$ expansion of the $[ij]$. In the first (cubic) term we can expand each factor to first order in $\log^{-1}\ep$ and the leading term will be proportional to $\la_1^2 \la_2^2 \la_3^2\,$, which makes a non-vanishing contribution under the $\la_i$ integrals. In the remaining (quadratic) terms however, we have to the expand each factor of $[ij]$ to second order in  $\log^{-1}\ep$, since the first order terms are odd in some $\la_i$ and thus vanish. Hence the contribution of these terms is subleading and of order $\log^{-4}\ep$.

It pays to again think of this in terms of graphs, except that now the edge factors are equal to $[ij]$. The first (cubic) term corresponds to a triangle (namely $\mathcal{C}_3$), and its leading contribution comes from the shortest loop\footnote{In graph theory this would be called a closed walk, since the word loop is reserved for an edge that connects a vertex to itself. There are no such loops in our graphs, so hopefully this abuse of terminology will not be confusing.} we can draw on a triangle (namely going around the triangle once) which corresponds to expanding each factor to first order. The quadratic terms on the other hand correspond to graphs with only two edges, and they contribute only once we expand every edge to second order, which again corresponds to the shortest loop we can draw on such a graph, namely going along the two edges and then back again. However this loop is of length four, and thus subleading compared to the triangle, which has a loop of length three.

This will be the general theme: we will write the result for $\langle q^K \rangle_c $ as a sum over terms labelled by certain graphs, and the contribution of each term will be suppressed by as many powers of $\log^{-1}\ep$ as the length shortest link (set of loops) we can draw on this graph.

As a slightly more non-trivial example, consider $K=4$. Again the expansion in terms of the $(ij)$ contains the complete graph $\mathcal{C}_4$ and all its complete subgraphs $\mathcal{C}_{4-n}$ with coefficient $(-1)^{4-n}$, accounting for multiplicities arising from the fact that we consider the vertices as labelled, i.e. distinguishable:
\[
&& (-64 \pi^4 \log\ep)^{2} \langle q^4 \rangle_c  \\
&& \ = (12)(13)(14)(23)(24)(34)  - (12)(13)(23) - (12)(14)(24) \nn \\
&& \ - (13)(14)(34) - (23)(24)(34) + (12) + (13) + (14) + (23) + (24) + (34) - 3  \nn \\ 
&& \ = [12][13][14][23][24][34]  \nn \\
&& \ + [13][14][23][24][34]  + (\mathrm{five\ similar\ terms\ with\ five\ factors}) \nn \\
&& \ + [14][23][24][34]  + (\mathrm{fourteen\ similar\ terms\ with\ four\ factors})  \nn \\
&& \ + [14][24][34]  + (\mathrm{fifteen\ similar\ terms\ with\ three\ factors\ and\ without\ closed\ loops}) \nn \\
&& \ + [12][34] + [13][24] + [14][23] \nn \ .
\]
The $[ij]$ expansion on the right hand side has a more interesting structure. It contains $\mathcal{C}_{4}$ which has six edges, all of its subgraphs with five edges and also every subgraph with four edges. Out of the 20 possible subgraphs with three edges, there are 16 present, and the four that are missing are precisely the complete graphs $\mathcal{C}_{3}$, i.e. triangles. Similarly, all subgraphs of these four $\mathcal{C}_{3}$ are absent and thus there no graphs with one or zero edges. The only graphs with two edges are the three that are not subgraphs of any of the $\mathcal{C}_{3}$. All those that appear do so with unit coefficient. 

In other words, the right hand side contains exactly one term for every subgraph of $\mathcal{C}_{4}$ which has at least one edge attached to every vertex (of $\mathcal{C}_{4}$), i.e. has no isolated vertices. 

Which terms give the leading contribution in this case? It is clear that the shortest link we can draw on a graph with $K$ vertices none of which are isolated, in such a way that every edge is traversed at least once, is of length $L_{\mathrm{tot}}=K$. In particular, if the expansion above contained $\mathcal{C}_{3}$, there would be loops of length $L=3$ covering that graph, but since this is not the case we have to look for loops of length $L=4$. Among the 15 graphs with four edges there are three containing a loop of length four. Furthermore, the last line contains three graphs which consists of only two edges. Each of those edges is completely disconnected from the rest of the graph, and we will refer to these as dimers. On a dimer we can draw a loop of length $L=2$ by going back and forth along the same edge, and we will call such loops trivial. Thus the three graphs in question also support links of length $L_{\mathrm{tot}}=4$, namely two mutually disconnected trivial loops each. Altogether only these six graphs, shown in Figure~1, will contribute at the leading order $\O(\log^{-4}\ep)$, which makes it much simpler to extract the dominant terms in the fourth central moment.

\begin{figure}
\begin{center}
\vspace{-1cm}
\includegraphics[width = 0.8 \textwidth]{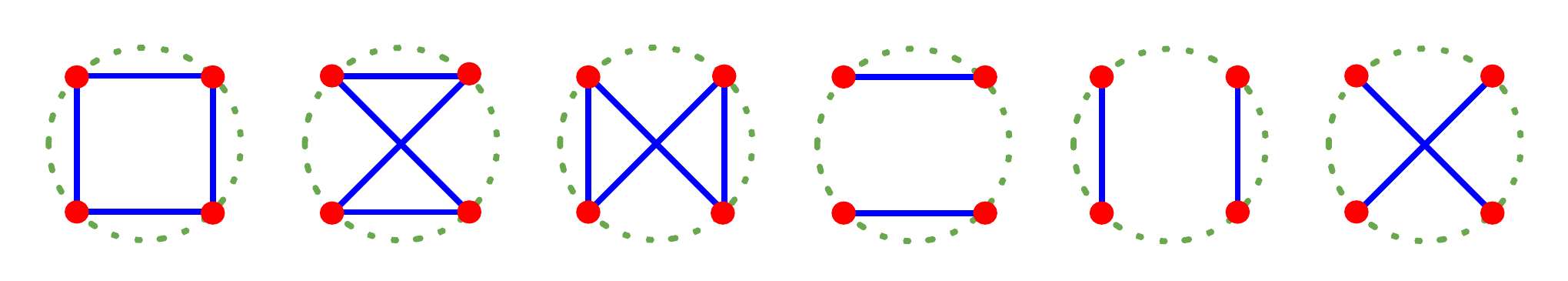} 
\vspace{-0.5cm}
\caption{\small The six graphs that contribute to the fourth central moment at leading order. For illustrative purposes we have chosen to arrange the four (distinguishable) vertices equidistantly on a circle, which can be thought of a constant time slice of $dS_2$.} 
\vspace{-0.5cm}
\end{center}
\end{figure}

For general $K$ we can easily write down the $(ij)$ expansion. It turns out that the overall factor we need to extract is equal to $(-64 \pi^4 \log\ep)^{-K/2}$. Thus we can write $\langle q^K \rangle_c$ as a sum over terms labelled by complete subgraphs of  $\mathcal{C}_{K}$, including singleton and null graphs.
\[ \label{CombRound}
(-64 \pi^4 \log\ep)^{K/2} \langle q^K \rangle_c  
= \sum_{\mathrm{complete\, graphs\, with}\,  n\, \mathrm{vertices}\, \subset \,\mathcal{C}_{K} } (-1)^{K-n} \prod_{j>i\ge1}^n (ij) \ . \qquad
\]

Our proposal is that when expanded in terms of the $[ij]$ this is equal to the much more useful expression
\[ \label{CombLemma}
(-64 \pi^4 \log\ep)^{K/2} \langle q^K \rangle_c  
= \sum_{\mathrm{graphs}\, \mathcal{G} \mathrm{with}\,  K\, \mathrm{non-isolated\, vertices}\, \subset \,\mathcal{C}_{K} } \ \ \prod_{\mathrm{edges}\, \overline{ij}\, \in\, \mathcal{G}} [ij] \ . \qquad
\]

We have checked explicitly that this holds also for $K=5$. The equality of the two graphical expansions on the right hand sides of these equations should be a simple consequence of the relation $[ij] \equiv (ij)-1$ (independent of the precise nature of the edge factor).\footnote{An iterative proof of this lemma for general $K$ might proceed along the following lines:  

First one considers just the term corresponding to the complete graph with $K$ vertices, i.e.~the product of all ${1 \over 2} K (K-1)$ round bracket edge factors $\prod_{\overline{ij}\, \in\, \mathcal{C}_{K} } (ij)$, and expands this in terms of rectangular brackets. It is not hard to see that this expansion contains a term $\prod_{\overline{ij}\, \in\, \mathcal{G}} [ij]$ with unit coefficient for every subgraph $\mathcal{G}$ of the completely graph $\mathcal{C}_K$, including the complete graph itself. This is because multiplying out products of the $(ij)-1$, in order to express everything in terms of round brackets, and summing over all subgraphs, only the desired term associated with the complete graph survives. All other terms come with vanishing coefficient, since for terms with $r$ edges removed from a total of  ${1 \over 2} K (K-1)$ we have $\sum_{p=0}^r (-1)^p {{1 \over 2} K (K-1) \choose p}{{1 \over 2} K (K-1) -p \choose r-p} = 0$.    

Now in the round bracket expansion (\ref{CombRound}) we are interested in, terms corresponding to the $K$ complete (sub)graphs with $K-1$ vertices are subtracted from this.  We can use the same identity again for each of them to express this as a rectangular bracket expansion containing terms with unit coefficient for every subgraph of  $\mathcal{C}_{K-1}$. Subtracting these corresponds to removing all terms that have exactly one isolated vertex (namely the one not contained in $\mathcal{C}_{K-1}$) from the rectangular expansion.

However, terms that have two isolated vertices have now been subtracted twice, so we compensate for this we add to the round bracket expansion the complete subgraphs $\mathcal{C}_{K-2}$ that do not contain these vertices, and so on until we get down to the singleton and null graphs. We end up with a round bracket expansion that is an alternating sum of complete (sub)graphs versus a rectangular bracket expansion that contains all subgraphs except those with isolated vertices.}
After all, it just an algebraic statement that two polynomials are equal, and the graphs simply serve as a convenient way of labeling the terms in the two expansions. 

\subsection{The Contribution of Cycle Graphs}\label{CycleGr}

Why is the latter expression (\ref{CombLemma}) so much more useful? We have already seen that it makes it obvious that the $K$th central moment will in fact be $\O(\log^{-3K/2}\ep)$, and allows us to easily identify a small subset of terms that contribute to the leading order. This is because the only graphs that contribute at this order are those that can be covered by links of length $L_{\mathrm{tot}}=K$ (i.e.~on which we can draw a set of loops using every edge of the graph at least once such that in total the number of steps does not exceed the number of vertices). It is easy to convince oneself that the relevant graphs are always composed of a set of (mutually disconnected) cycle graphs (in which every vertex has degree two) and dimers (two vertices of degree one joined by a single edge).

Not only are these graphs easy to identify, but we can also calculate their contribution explicitly. For a cycle graph this is much simpler than for a large connected graph (which we would have to compute in order to find the $K$th moment directly), since every vertex is connected only to two instead of $K-1$ other vertices.
  
Using the Fourier expansion of the logarithm of $(ij)$ given in (\ref{FourierEdge}) we find
\[
&&\int_0^{2\pi} {\d \th_i \over 2 \pi} \left[\left(4 \sin^2 {\th_{i-1}-\th_i \over 2}\right)^{-\la_{i-1}\la_i/\log{\ep}} - 1\right] \left[\left(4 \sin^2 {\th_{i}-\th_{i+1} \over 2}\right)^{-\la_{i}\la_{i+1}/\log{\ep}} - 1\right] \nn \\
&=& {\la_{i-1} \la_i^2 \la_{i+1} \over \log^2 \ep} \sum_{n=1}^\infty {2 \over n^2} \cos(n(\th_{i-1} - \th_{i+1})) + \O(\log^{-3}\ep) \ .
\]
Concatenating further edge factors with this expression is trivial, since the result is self-similar - it simply leads to further factors of $\la_{j}\la_{j+1} \log^{-1} \ep$ and increases the power of $n$ in the denominator that multiplies the cosine of the difference between the first and the last angle. Once we close the loop after $L$ steps, there will be precisely two powers of $\la_i$ for every vertex we passed through, and the sum of $n^{-L}$ will result in a zeta value $\zeta(L)$. Thus if we identify $\th_L = \th_0$ and $\la_L = \la_0$ we find for $L \ge 3$
\[ \label{ZetaValue}
\prod_{i=1}^L [i-1,i] 
&=& \int_{-\infty}^{\infty} \left(\prod_{i=1}^L {\d\la_i \over \sqrt{\pi}}\, e^{-\la_i^2}\right) \int_0^{2\pi}  \left(\prod_{j=1}^L {\d\th_j \over 2\pi} \right)\, \prod_{k=1}^L \left[\left(4 \sin^2 {\th_{k-1}-\th_k \over 2}\right)^{-\la_{k-1}\la_k/\log{\ep}} - 1\right] \nn\\
&=& (2 \log \ep)^{-L}\, 2\, \zeta(L) + \O(\log^{-L-2} \ep)\ . 
\]

For trivial loops on dimers ($L=2$) the same result holds, except that there is an extra factor of $1/2$ since in this case we use the same edge twice and therefore have to expand to second order.
  
Thus a loop of length $L$ contributes a factor proportional to $\zeta(L)$. For a graph consisting of a number of disconnected cycles and dimers we simply multiply the contribution from each component. Since the total length of all loops in the link has to add up to $K$, this nicely confirms that the leading term of the (rescaled) $K$th central moment will have transcendentality $K$.

\subsection{Counting Graphs}

We have identified the graphs that contribute to the leading term of the central moments, and calculated the individual contributions of such graphs. What remains to be done is to count how many graphs will contribute for a given $K$. We will do this in two steps. First we need to enumerate the different families of graphs that are relevant, and then count the number of graphs within each given family.

We distinguish families simply according to how many (mutually disconnected) loops of length $L$ are present in the graph. We can characterize this by a set of integers $m_L$ that give the multiplicities of loops of a given length. In particular, $m_2$ gives the number of dimers, and the $m_L$ for $L>2$ count the number of cycles of length $L$ contained in the graph. We have already argued above that the graphs relevant to the leading central moments must obey the constraint
\[
\sum_{L \ge 2}\, m_L L = K \ .
\]

In other words, the families of graphs relevant for the $K$th central moment are labelled by the integer partitions of $K$ with the additional constraint that each summand has to be at least two.\footnote{This is known as an intermediate function in number theory.} We will denote the space of such integer partitions by $\mathsf{P}(2 , K)$. 

How many graphs are there in a given family? If there are $m_L$ cycles of a given length $L$ we can begin by choosing $m_L$ indistinguishable sets of $L$ out of $K$ vertices, which can be done in
\[
{1 \over m_L!} {K \choose L} {K-L \choose L} \ldots {K - (m_L-1)L \choose L} = {K! \over m_L!\, (L!)^{m_L}\, (K- m_L L)! }
\] 
ways. E.g.~we can choose to start with $L=2$ and then pick $m_3$ indistinguishable sets of three vertices out of the remaining $K - 2 m_2$ vertices and so on, which results simply in a product of factors of the form given above with $K$ replaced by the number of vertices remaining after each step.

Once we have picked sets of vertices we have to count how many different ways there are to make a cycle out of $L$ vertices. It easy to see that this is $L!$ modulo discrete rotations and orientation reversal, i.e. $(L-1)!/2$.

Again, the $L=2$ dimer is a special case, since there is clearly exactly one way of connecting to vertices with one edge, i.e.~the formula for cycles must be amended by an extra factor of 2. 

In summary, for a given integer partition $m_L \in \mathsf{P}(2 , K)$, which specifies the family, the number of different graphs is given by
\[ \label{GraphCount}
K!\, 2^{m_2} \prod_{L \ge 2} {1 \over m_L! (2L)^{m_L}} \ .
\]

\subsection{Explicit Expressions for Central Moments}

Putting everything together, the leading central moments that follow from (\ref{CombLemma}) by summing over all families of graphs labelled by integer partitions, making contributions (\ref{ZetaValue}) with multiplicities (\ref{GraphCount}), are given by
\[ \label{CentralMoments}
\langle q^K \rangle_c =  (-64 \pi^4 \log\ep)^{-K/2} (2 \log \ep)^{-K} K! \, \sum_{m_L \in \mathsf{P}(2 , K)}\,  \prod_{L \ge 2} {1 \over m_L! } \left({\zeta(L) \over L}\right)^{m_L} + \O(\log^{-3K/2-1}\ep) \ . \qquad
\]
Happily, the extra factors of $1/2$ from having to expand to second order for dimers precisely cancel the extra factors of 2 from counting graphs with dimers, so that the $L=2$ case appears no different than $L \ge 3$ in the end.

This expression agrees with the leading terms in $\langle q^2 \rangle_c$ and $\langle q^3 \rangle_c$ found above, and gives the following results for the subsequent central moments
\[
\langle q^4 \rangle_c &=& (-256 \pi^4 \log^3 \ep)^{-2}\  {3 \pi^4 \over 20} + \O(\log^{-7} \ep) \ , \\
\langle q^5 \rangle_c &=& (-256 \pi^4 \log^3 \ep)^{-5/2}\  \left({10 \pi^2 \over 3} \zeta(3) + 24 \zeta(5) \right) + \O(\log^{-17/2} \ep) \ , \\
\langle q^6 \rangle_c &=& (-256 \pi^4 \log^3 \ep)^{-3}\  \left({61 \pi^6 \over 168} + 40 \zeta(3)^2 \right)  + \O(\log^{-10} \ep) \ ,
\]
and so on. Here we have made use of the fact that zeta functions of even argument are expressible as rational numbers times powers of $\pi$. 
As remarked above the central moments have uniform transcendentality and are $\O(\log^{-3K/2} \ep)$. This motivates us to define a new variable
\[
\hat{q} \equiv (-256 \pi^4 \log^3 \ep)^{1/2} (q - \langle q \rangle) \ , 
\]
in terms of which the overlap distribution will be smooth in the small $\ep$ limit.

\subsection{Characteristic Function and Overlap Distribution}

Knowing all central moments (at least to leading order) we would now like to reconstruct the overlap distribution. This is best done via the characteristic function, which is the expectation value of a complex exponential with frequency $\omega$.

Our result (\ref{CentralMoments}) involves a sum over integer partitions, so it is helpful to know that the generating function of (unrestricted) partitions is given by $\prod_{L = 1}^\infty (1-x^L)^{-1}$. We can easily modify this for our restricted case in which each summand has to be at least two, by simply omitting the $L=1$ factor. However, this is still not exactly what we need, since we are interested in integer partitions that come with additional factors of $1/m_L!$ depending on the multiplicities with which the integers $L$ appear. Fortunately for us, this case is even easier: the generating function is simply a product of exponentials, and thus
\[ \label{character}
\left\langle e^{i \omega \hat{q}} \right\rangle &=& \sum_{K=0}^\infty \langle\, \hat{q}^K \rangle\, { (i \omega)^K \over K! } = \prod_{L \ge 2} \exp\left((i \omega)^L {\zeta(L) \over L}\right)  \nn \\
&=& e^{-i \gamma \omega}\ \Gamma(1 - i \omega) \ ,
\]
where we have used $\sum_{L \ge 2} (i \omega)^L \zeta(L)/L = - i \gamma \omega + \log(\Gamma(1 - i \omega))$ and $\gamma$ is the Euler-Mascheroni constant.

Using the integral representation of the $\Gamma$-function we can then Fourier transform the characteristic function to obtain the original overlap distribution for the (shifted and rescaled) variable $\hat{q}$:
\[ \label{Gumbel}
\mathcal{P}(\hat{q}) &=& {1 \over 2 \pi} \int_{-\infty}^{\infty}  \d \omega \, e^{-i \omega \hat{q}} \left\langle e^{i \omega \hat{q}} \right\rangle = \int_0^\infty \d t \, e^{-t}\, \delta(\hat{q} + \gamma + \log{t}) \nn \\
&=& \exp(-\hat{q}-\gamma-e^{-\hat{q}-\gamma}) \ .
\]

This is known as a Gumbel or log-Weibull distribution (see Figure~3 below). It is easy to check that it is correctly normalized and reproduces the leading terms of all the central moments we have computed above. 

Of course we have assumed here that the domain of $\mathcal{P}(\hat{q})$ is the whole real line. While this is not strictly true, it is not a problem in the small $\ep$ limit we are interested in, since $\hat{q}$ is a rescaled variable, and in terms of the original variable $q$ the distribution looks highly compressed (almost like a $\delta$-function). The distribution falls off exponentially in the positive direction, and even faster (as a double exponential) in the negative direction, so any tails outside the original domain of $\mathcal{P}(q)$ are highly suppressed for small $\ep$.

\section{Triple Overlap} \label{Triple}

We will now turn to a finer probe of the structure of the space of states, namely the triple overlap distribution. Here we consider three copies (replicas) of the system under consideration, and compute the multivariate distribution of mutual overlaps $q_{ab}$ between them.
In order to avoid an overabundance of indices we will use dual variables $q_a \equiv |\varepsilon_{abc}|\, q_{bc}/2$, and similarly for other quantities.

\subsection{Moments in Terms of Wigner Functions}

We can express the moments of this distribution in terms of coarse-grained Wigner functions as follows:     
\[
\langle q_{1}^{K_{1}} q_{2}^{K_{2}}q_{3}^{K_{3}} \rangle &=& \int_0^{2\pi} \prod_{i=0}^K {\d \th_i \over 2 \pi} \int \prod_{j=1}^K \d X_j \d P_j \\
&\times& \left(W_1^{K_{1}+1, K_{1}+2, \ldots ,K} \right) \left(W_2^{1,2, \ldots, K_{1}, K_{1}+K_{2}+1, K_{1}+K_{2}+2 \ldots ,K} \right) 
\left(W_3^{1,2, \ldots, K_{1}+K_{2}} \right) \ ,  \nn
\] 
where there are three groups of degrees of freedom\footnote{Note that the $K_a$ are dual variables corresponding to the more natural $K_{ab}$, whereas the Wigner distributions have only one index to begin with since they are associated with one replica only.} such that $K = K_{1}+K_{2}+K_{3}$. For simplicity we have chosen a particular ordering of the $K$ degrees of freedom, which makes very explicit that every set of variables $(\th_i,X_i,P_i)$ for a given index $i$ is associated with exactly two of the three replicas (but of course any permutation would be just as good). Reconstructing the full triple overlap distribution from its moments will allow us to determine, amongst other features of the space of states, to what extent the system exhibits ultrametricity. 

We can use the representation (\ref{Wigner}) for the Wigner functionals, where the parameters $\la_k$ and $\tilde{\la}_k$ should be thought of as carrying an additional index to indicate which replica they belong to. As above, however, performing the integral over the coarse-grained expectation values $X_i$ and corresponding momenta $P_i$ leads to delta functions that allow us to eliminate all but one complete set of $(\la_k, \tilde{\la}_k)$ parameters, such that the extra index essentially disappears.

To illustrate this in more detail, let's assume that the momentum integral again only contributes terms suppressed by positive powers of $\ep$, and rescale the $\la_k$ as above, so that up to numerical factors the Wigner distribution effectively can be written as
\[
W_a^{1,2,3\ldots K} \sim \int_{-\infty}^\infty \prod_{i = 1}^K \left(\d \la_i^a\, e^{-(\la_i^a)^2/2}\, e^{2 \pi i \la_i^a X_i'}\right) \prod_{k>j\ge1}^K \left(4 \sin^2 { \th_k-\th_j \over 2}\right)^{-\la_j^a \la_k^a / (2\log\ep)} \ .
\]

Multiplying three such factors and integrating over the (for simplicity also rescaled) $X_i'$ will set sums of two $\la_i^a$ with the same $i$ but different replica index equal to zero. Integrating over those delta functions, and choosing the remaining set of parameters $\la_i$ in a symmetric fashion, the resulting formula for the moments of the triple overlap distribution reduces to basically the same expression (\ref{HigherInt}) we have already encountered.

The overall $K$ is now interpreted as the sum of the $K_a$, and the left-over parameters $\la_i$ are integrated over with Gaussian weights as before, except for one crucial difference: the angular factors connecting two degrees of freedom in different groups have an extra factor of $-{1 \over 2}$ in the exponent. This is clear, since as shown in Figure~2, those factors appear only in one of the three Wigner distributions (rather than in two of them such as those connecting degrees of freedom in the same group). The most obvious choice of parameters is to define e.g. $\la_i \equiv \la_i^2 = - \la_i^3$ for $i = 1,2,\ldots,K_1$, which introduces minus signs whenever these is a connection between the first and the second group. Similarity transformations, leading to less symmetric choices of parameters, can eliminate some, but not all of these minus signs. 

To put it differently, if we carry out the Gaussian $\la_i$ integrals first and think of the problem as computing the (inverse square root of the) determinant of a $K$ by $K$ matrix $M_{ij}$, the crucial difference in the structure of the relevant matrix compared to the $M_{ij}$ considered above is that block off-diagonal elements are multiplied by $-{1 \over 2}$, whereas entries in the three $K_a$ by $K_a$ blocks on the diagonal are the same as before. This is very reminiscent of the form of the overlap matrix in a system with one-step replica symmetry breaking (even though there appears to be no direct physical relation).

\begin{figure}
\begin{center}
\vspace{-1cm}
\includegraphics[width = 0.9 \textwidth]{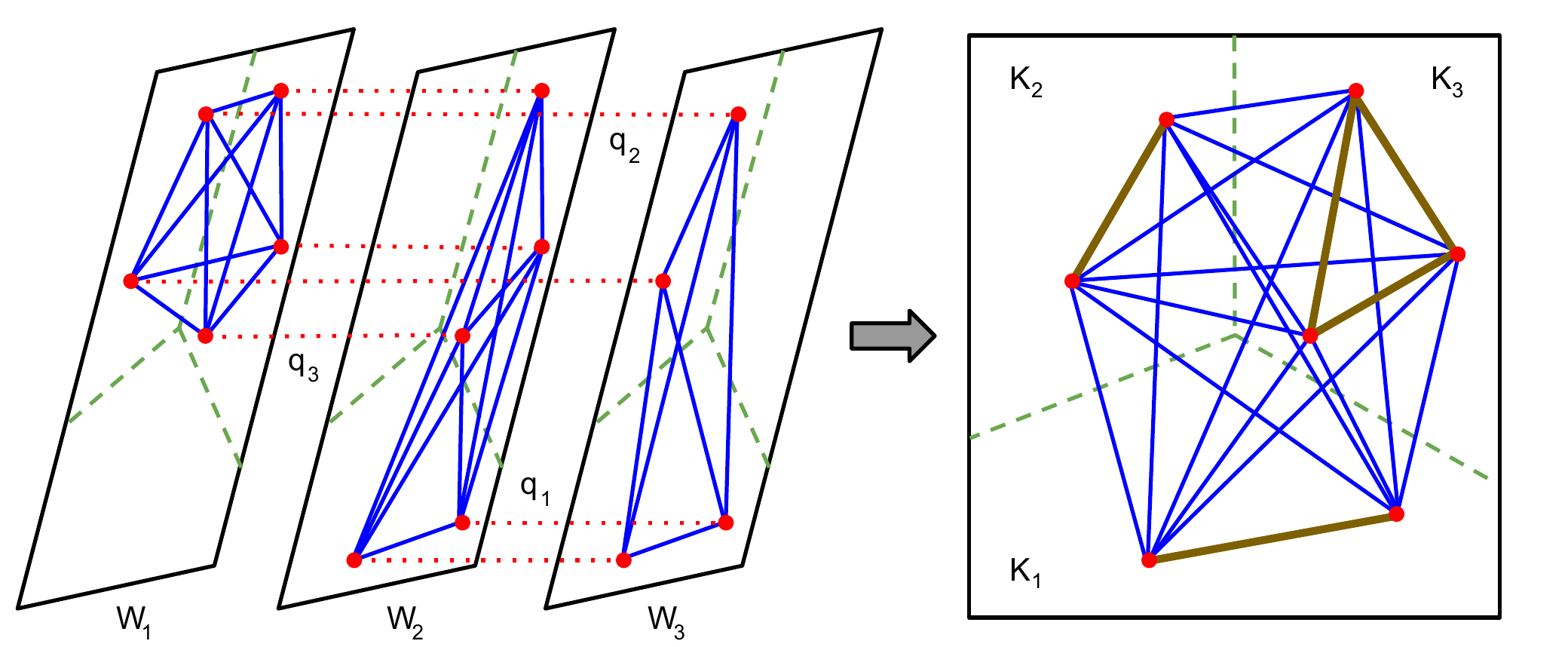} 
\vspace{-0.1cm}
\caption{\small Illustration of the triple overlap computation. On the left the three copies of the Wigner distribution are shown, with all edge factors that appear in them. One the right, we superimpose the three pictures, which corresponds to the situation after integrating over the coarse-grained variables $(X_i, P_i)$ and eliminating all $(\la_i,\tilde{\la}_i)$ parameters that appear in $\de$-functions. Edges within one group appear twice on the left, and after being identified are drawn as thick lines on the right. Edges that stretch between two groups appear only once, and are suppressed by a factor of $-{1 \over 2}$.} 
\vspace{-0.5cm}
\end{center}
\end{figure}

\subsection{Computing Moments Using Graphical Techniques}

Since the expressions for the moments of the distribution are so similar, the calculations of Section \ref{Low} translate immediately into results for the first few moments of the triple overlap distribution. This will not take us very far however, so we'll turn directly to applying the method of Section \ref{Graphs} to this problem.

We have learned that we should really be computing the central moments
\[
\langle \hat{q}_{1}^{K_{1}} \hat{q}_{2}^{K_{2}} \hat{q}_{3}^{K_{3}} \rangle \equiv (-256 \pi^4 \log^3 \ep)^{K/2} \left\langle (q_{1} - \langle q_1 \rangle)^{K_{1}} (q_{2}- \langle q_2 \rangle)^{K_{2}}(q_{3}- \langle q_3 \rangle)^{K_{3}} \right\rangle \ .
\]

As above, we will associate to each of these central moments a set of graphs, with $K$ vertices labelled by the index $i$, and edges corresponding to the angular factors connecting two vertices (i.e.~depending on the difference of two degrees of freedom). However, now there is additional structure to the problem: the vertices are split up into three mutually exclusive groups, with the number in each group given by $K_a$, and the edges come in two varieties, namely those connecting two vertices within one group, and those connecting different groups.

The arguments of Subsection \ref{EdgeExp} go through as before, since they don't rely on the exact form of the edge factors, and thus we can write the leading term of each central moment as a sum of terms labelled by graphs with $K$ non-isolated vertices on which minimal length links can be drawn.  

The contribution of each cycle or dimer contained within such a graph also follows easily from the calculation in Subsection \ref{CycleGr}. We simply need to insert appropriate factors of $-{1 \over 2}$ whenever we cross the boundary between different groups of vertices.

However, we face a rather nontrivial counting problem: how many (unoriented) graphs with $K$ vertices are there which consist of disconnected cycle or dimer components (subgraphs), in which each vertex is part of exactly one disconnect component, and also belongs to exactly one of three distinguishable groups with $K_a$ members, such that the boundaries between groups are crossed a given number of times?

\subsection{More Graph Combinatorics}

Of course we are not really interested in the number of graphs as such, but rather would like to compute a weighted sum over all of them with the appropriate factors attached to each one, and thus we need a complete classification of the graphs that contribute to the central moments at leading order. We will do this in several steps. 

At the highest level, the relevant graphs are characterized by the family they belong to. As discussed in the previous section, we specify the family by an integer partition of K written as $m_L \in \mathsf{P}(2 , K)$ (which each summand at least two), which counts the multiplicities of loops (or dimers) of length $L$. We know from the above that within each family there are $K!\, 2^{m_2} \prod_{L \ge 2} (m_L!)^{-1} (2L)^{-m_L}$ distinct graphs. We obtained this by first counting the distinguishable ways of assigning each of the $K$ vertices to one of the loops, and then multiplying by the number of ways we can string together (wire up) a given set of $L$ vertices into a loop of length $L$. 

For the present problem however, we need a finer classification, since vertices are now assigned to one of three groups and we need to consider whether edges cross between different groups. Therefore, when assigning vertices to loops of a certain length $L$ we have to consider how many vertices to take from each of the three groups. The number of ways of doing this is given by
\[
&&{1 \over m_L!} \sum_{n_1^1,n_2^1,n_3^1 = 0}{K_1 \choose n_1^1} {K_2 \choose n_2^1} {K_3 \choose n_3^1} \delta(n_1^1+n_2^1+n_3^1-L) \times \ldots \nn \\
&\times& \sum_{n_1^{m_L},n_2^{m_L},n_3^{m_L} = 0}{K_1 \choose n_1^{m_L}} {K_2 \choose n_2^{m_L}} {K_3 \choose n_3^{m_L}} \delta(n_1^{m_L}+n_2^{m_L}+n_3^{m_L}-L) \nn \\
&=&{1 \over m_L!} \sum_{n_1^1\ldots n_1^{m_L} = 0} \sum_{n_2^1\ldots n_2^{m_L} = 0} {K_1!K_2!K_3! \over n^1_1!\, n^1_2!\, (L-n^1_1-n^1_2)!\ldots n^{m_L}_1!\,n^{m_L}_2!\,(L-n^{m_L}_1-n^{m_L}_2)! }  \nn \\
&\times& {1 \over (K_1 - \sum_{\mu=1}^{m_L} n_1^\mu)!(K_2 - \sum_{\nu=1}^{m_L} n_2^\nu)!(K_3 - \sum_{\rho=1}^{m_L} (L-n_1^\rho-n_2^\rho))!} \ .
\] 

The factor of $1/m_L!$ appears because loops of the same length are interchangeable at this level. For each loop there is obviously a constraint that the numbers of the vertices drawn from the three groups have to add up to $L$. When we go through all the possible values of $L$ we will multiply factors of this type until all vertices are assigned. Thus, when we are given a set of integers $\{K_1,K_2,K_3\}$ there are also global constraints which impose that the total number of vertices drawn from each group has to add up to the correct $K_a$. Demanding this, the total number of possible vertex assignments is 
\[
\sum_{\{n_1^{L,\mu_L}\} = 0}^\infty\,  \sum_{\{n_2^{L,\mu_L}\} = 0}^\infty && K_1! K_2! K_3!\, \de(K_1 - \sum_{L, \nu_L} n_1^{L,\nu_L})\, \de(K_2 - \sum_{L, \nu_L} n_2^{L,\nu_L}) \nn \\
&& \prod_{L \ge 2} {1 \over m_L!} \prod_{\mu_L=1}^{m_L} \,{1 \over n^{L, \mu_L}_1!\,n^{L, \mu_L}_2!\,(L-n^{L, \mu_L}_1-n^{L, \mu_L}_2)!} \ ,
\]
where there are parameters $n^{L,\mu_L}$ appearing inside the sums (and products) for every $L$ for which $m_L >0$ and $\mu_L = 1,2 \ldots m_L$.

Thus for given $K_a$ we can define a genus (think biology, not topology) of graphs belonging to a given family $m_L$ by a set of pairs of integers $\{n_1^{L,\mu_L}, n_2^{L,\mu_L} \}$  where pairs associated with the same length $L$ but different labels $\mu_L$ are interchangeable (i.e.~sets differing only by swapping pairs with the same $L$ are considered equivalent). A valid set has to satisfy the local constraints $n_1^{L,\mu_L} + n_2^{L,\mu_L} \le L$ for all $L$ and $\mu_L$ and the global constraints $\sum_{L,\mu_L} n_a^{L,\mu_L} = K_a$.

However, even within a given genus there are graphs with different factors associated with them, because even once we have assigned vertices to loops of given length there are still different ways of wiring up the vertices into a cycle, which may cross the boundaries between groups a different number of times. As a simple example, consider a cycle of length four for which we have chosen two vertices from one group and two vertices from another. Clearly, we can draw loops of length four that cross between the groups either twice or four times (as is apparent from the three leftmost graphs in Figure~1, considering say the top two vertices as belonging to a different group than the bottom two). 

Thus within a given family and genus, we can think of a species of graphs as a subset which for each cycle it contains has a specified number of edges between vertices belonging to different groups, each of which picks up a factor if $-{1 \over 2}$ relative to edges within one group. We define for each cycle a loop factor $\mathcal{L}^L(n_1, n_2)$ that sums over all possible cycle graphs within a given genus and weighs each species by the appropriate power of $-{1 \over 2}$ (taking into account the correct population of each species). In the example of the previous paragraph, we can draw two different graphs that cross boundaries between groups twice and one graph that has four crossings. Therefore for this loop the correct factor would be $\mathcal{L}^4(n_1=2, n_2=2) = 2 (-{1 \over 2})^2 + 1 (-{1 \over 2})^4 = {9 \over 16}$.
The total loop factor will simply be a product of such factors for each cycle (or dimer) contained in the graph.

In summary, in order to compute a given moment with exponents $K_a$ all we need to do is sum over all relevant families, genera and species of graphs with zeta function values depending on the family, and loop factors taking care of the numerical suppression associated with edges crossing borders  as well as multiplicities within each species. To execute this we would have to overcome two difficulties: firstly we would need an explicit formula for the loop factors $\mathcal{L}^L(n_1, n_2)$, and secondly we would have to find an efficient way of implementing the global constraints when summing over genera. Fortunately, it is not necessary to perform either of these steps explicitly.

\subsection{Characteristic Function from Matrix Powers}

An elegant way to avoid the ugly details that arise when computing individual moments in this fashion is to work with generating functions. Consider the matrix
\[
\Omega = \left( \begin{array}{ccc}
\om_1 & 0 & 0 \\
0 & \om_2 & 0 \\
0 & 0 & \om_3 \end{array} \right)
\left( \begin{array}{ccc}
1 & -1/2 & -1/2 \\
-1/2 & 1 & -1/2 \\
-1/2 & -1/2 & 1 \end{array} \right) = 
\left( \begin{array}{ccc}
\om_1 & -\om_1/2 & -\om_1/2 \\
-\om_2/2 & \om_2 & -\om_2/2 \\
-\om_3/2 & -\om_3/2 & \om_3 \end{array} \right) \ .
\]
It turns out that it generates precisely the loop factors we require
\[ \label{Omega}
{1 + \de_{L,2} \over 2L}\, \mathrm{tr}(\Omega^L) =\sum_{n_1+n_2+n_3 = L} {\mathcal{L}^L(n_1, n_2) \over n_1!\, n_2!\, n_3!}\  \om_1^{n_1}\om_2^{n_2}\om_3^{n_3} \ .
\]
Intuitively, the matrix powers of $\Omega$ encode the fact that there is a penalty for going from on group to another (in the off-diagonal elements of $\Omega$), while at the same time taking into account all the different paths that visit each vertex once. The trace makes sure that the path closes in the end. Again, the dimer case is special and gives rise to an extra factor of two.

The above expression, which gives us a linear combination of all loop factors for a given $L$, rather than just a particular one, is precisely what we need to compute the characteristic function of the triple overlap distribution. This also avoids the second problem of having to impose global constraints: since we have to sum over all $K_a$, the issue no longer arises.  

Neglecting all subleading terms (suppressed by additional inverse powers of $\log\ep$) the characteristic function is then given by
\[
&&\langle \exp(i \om_1 \hat{q}_1 + i \om_2 \hat{q}_2 + i \om_3 \hat{q}_3)\rangle =  \sum_{K_1, K_2, K_3 = 0}^\infty {(i \om_1)^{K_1}(i \om_2)^{K_2}(i \om_3)^{K_3} \over K_1!\, K_2!\, K_3!} \langle \hat{q}_{1}^{K_{1}} \hat{q}_{2}^{K_{2}} \hat{q}_{3}^{K_{3}} \rangle \nn \\
&=&  \sum_{K_1, K_2, K_3 = 0}^\infty \sum_{m_L \in \mathsf{P}(2 , K)} \sum_{\{n_1^{L,\mu_L}\} = 0}^\infty \sum_{\{n_2^{L,\mu_L}\} = 0}^\infty  \sum_{\{n_3^{L,\mu_L}\} = 0}^\infty \left(\prod_{a=1}^3 \de(K_a - \sum_{L, \nu_L} n_a^{L,\nu_L}) \right) \nn \\
&&\times (i \om_1)^{K_1}(i \om_2)^{K_2}(i \om_3)^{K_3} \prod_{L \ge 2} {1 \over m_L!} \prod_{\mu_L =1}^{m_L}  {\mathcal{L}^L(n_1, n_2) \over n_1!\, n_2!\, n_3!} \, \de(n_1^{L,\mu_L}+n_2^{L,\mu_L}+n_3^{L,\mu_L}-L) \ 2 \zeta(L) \ . \nn \\
&=&   \exp\left(\sum_{L \ge 2} {i^L \zeta(L) \over L} \, \mathrm{tr}(\Omega^L) \right) \ .
\]
Again, the extra factors of two for the dimer cancel, and performing the sums over the $K_a$ first, then summing over the $n_a^{L,\mu_L}$ using (\ref{Omega}) and finally using the generating function for integer partitions, the expression simplifies dramatically to a non-Abelian version of (\ref{character}).

\subsection{A Tale of Two Gumbels}

The non-zero eigenvalues of $\Omega$ are given by 
\[
\om_\pm = {1\over 2}(\om_1+\om_2+\om_3)  \pm {1 \over 2 } \sqrt{\om_1^2+\om_2^2+\om_3^2-\om_1\om_2-\om_1\om_3-\om_2\om_3} \ ,
\]
and therefore the trace of $\Omega^L$ is simply equal to $\om_+^L + \om_-^L$.

If we introduce cylindrical polar coordinates aligned with the equilateral axis
\[
\om_z &=& {1 \over \sqrt{3}} (\om_1+\om_2+\om_3) \ , \nn \\
\om_r &=& \sqrt{2 \over 3} \sqrt{\om_1^2+\om_2^2+\om_3^2-\om_1\om_2-\om_1\om_3-\om_2\om_3} \ , \nn \\
\om_\varphi &=& \arctan\left({\sqrt{3}(\om_1 - \om_3) \over \om_1 - 2 \om_2 + \om_3}\right) \ ,
\]
and similarly for the dual coordinates $(\hat{q}_z,\hat{q}_r,\hat{q}_\varphi)$, we find that the characteristic function can be written as
\[ \label{TODChar}
\left\langle e^{i \om_z \hat{q}_z + i \om_r \hat{q}_r \cos(\om_\varphi - \hat{q}_\varphi) } \right\rangle = e^{- i \gamma \sqrt{3}\, \om_z} \Gamma(1 - i \om_+) \Gamma(1 - i \om_-) \ ,
\]
where we have performed the sum over $L$ explicitly, and
$\om_\pm =  (\sqrt{3}/ 2) ( \om_z  \pm  \om_r/\sqrt{2})$.

The characteristic function of the triple overlap distribution is independent of the angular variable $\omega_\varphi$. In fact, it is simply a product of the characteristic functions of two Gumbel distributions, and thus its Fourier transform will lead to an expression resembling  a convolution (though not exactly, because the $\om_\pm$ integration region is only a half-plane). Performing this Fourier transform we obtain, using again the integral representation of the $\Gamma$-function
\[
\mathcal{P}(\hat{q}_1, \hat{q}_2, \hat{q}_3) &=& {1 \over (2 \pi)^3} \int \d \om_z \d \om_r \d \om_\varphi \ \om_r \, e^{-i \om.\hat{q}} \, e^{-i \sqrt{3} \gamma \om_z} \\ 
&& \int_0^\infty \d s \,  e^{-s} \int_0^\infty \d t \,  e^{-t} \, e^{-i \sqrt{3}(\log s + \log t)\, \om_z  / 2 }  \, e^{-i \sqrt{3}(\log s - \log t)\, \om_r  / (2 \sqrt{2}) } \ . \nn
\]

Since the integrand is independent of $\om_\varphi$ except for the complex exponential of $\om.\hat{q}$, the $\om_\varphi$ integral will simply pick out the lowest order Bessel function of the angular expansion of this factor, and the result will be independent of $\hat{q}_\varphi$. The $\om_z$ integral leads to a delta function which trivializes one of the parametric integrals. After an appropriate change of variables, the second parametric integral then takes the form of the integral representation of the modified Bessel function of the second kind
\[
K_\nu(x) = \int_0^\infty \d u \, e^{-x \cosh u} \cosh(\nu u) \ ,
\]
but with purely imaginary order $\nu$. This leads to the following expression for the triple overlap distribution
\[ \label{TripleDis}
\mathcal{P}(\hat{q}_z, \hat{q}_r) = {2 \over \pi \sqrt{3}} \int_0^\infty \d \om_r \, \om_r \, J_0(\om_r \hat{q}_r) \, e^{-2(\gamma + \hat{q}_z / \sqrt{3})} K_{i \sqrt{3}\, \om_r / \sqrt{2}} \left(2 e^{-(\gamma + \hat{q}_z / \sqrt{3})} \right) \ . 
\]
This peculiar type of integral, with the integration variable appearing in the order of the modified Bessel function, is known as a Kontorovich-Lebedev transform (or its inverse, depending on conventions).
One can check that this probability density is correctly normalized, and that the first moment $\langle \hat{q}_z \rangle$ vanishes\footnote{This might not be apparent from the plot of the probability density, which for small $\hat{q}_r$ is clearly peaked in a region of negative $\hat{q}_z$. Recall however that the integration measure contains another factor of $\hat{q}_r$, which mitigates this, and enhances the tail of the distribution which is more pronounced for positive $\hat{q}_z$.} as required, though for the computation of higher moments it is more convenient to directly expand the characteristic function (\ref{TODChar}) instead.

\begin{figure}
\begin{center}
\vspace{-2.8cm}
\includegraphics[width = 0.43 \textwidth]{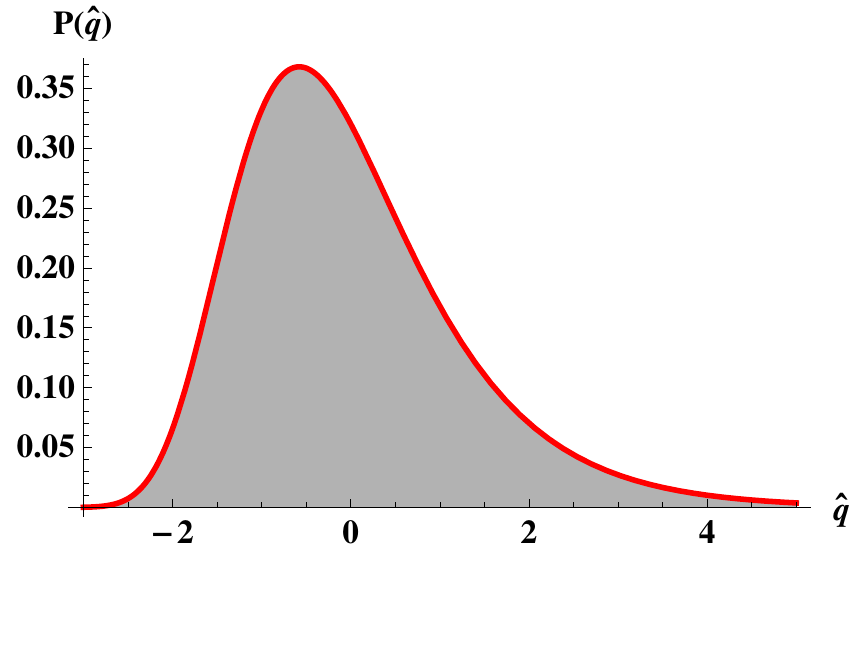}
\hspace{-1cm}
\includegraphics[width = 0.6 \textwidth]{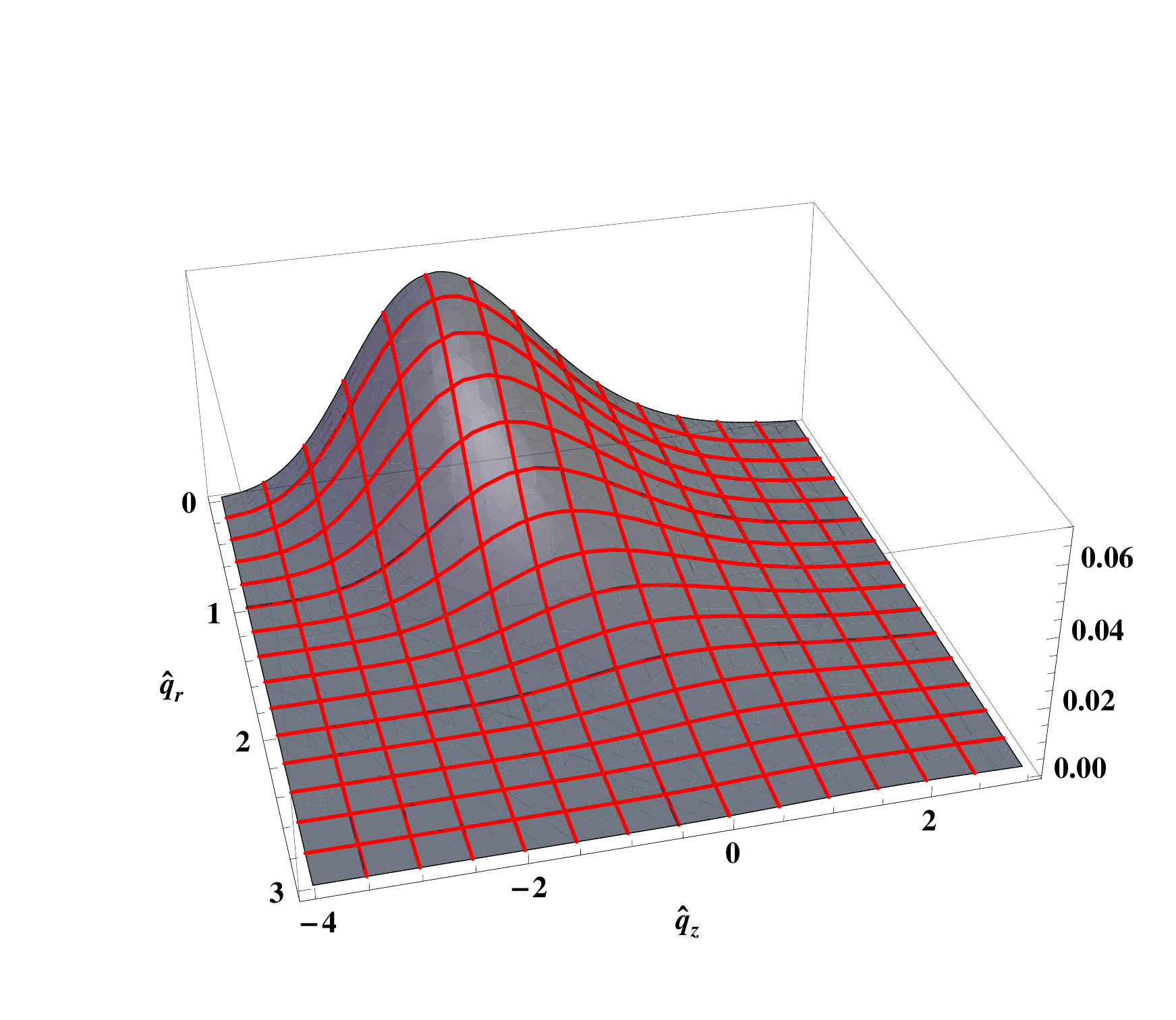} 
\vspace{-1cm}
\caption{\small Overlap distribution $\mathcal{P}(\hat{q})$ (left), and triple overlap distribution $\mathcal{P}(\hat{q}_z, \hat{q}_r)$ (right).} \vspace{-0.5cm}
\end{center}
\end{figure}

\section{Discussion}

We have introduced a neat graph-based technique that allowed us to explicitly compute the overlap and triple overlap distributions for a coarse-grained, massless scalar field on a 1+1 dimensional de Sitter background by largely combinatorial methods.

What have we learned from these calculations? The obvious answer, which was of course entirely expected, is that in the naive late time $\ep \to 0$ limit, in which the number of causally disconnected boxes in our model universe diverges, the overlap of a scalar field in $dS_2$ tends to vanish (i.e.~the limiting distribution is a delta function centered on zero).
This simply  confirms our intuition that in this limit the convolution with the box function merely picks out a perfectly localized harmonic oscillator degree of freedom, and since the underlying field theory is free there is no reason to expect anything non-trivial to emerge from such a collection of harmonic oscillators. 

However, it is a remarkable fact that if we look more closely, subtract the mean and scale the central moments of the overlap distribution appropriately, we find a rather non-trivial shape for the overlap, which is given by a Gumbel distribution (\ref{Gumbel}). Thus we have obtained a precise characterization of how likely deviations from the above trivial behavior are. 
Whether we accept this as evidence for a non-trivial phase structure, depends primarily on whether or not we consider the rescaled distribution physically relevant. 

There may be good, physical reasons to perform this shift and rescaling, since in a sense it merely removes the obvious effect of an inflating universe, which is the tendency to dilute overlaps between states on an absolute scale, maybe not unlike the situation in a spin glass to which we keep adding spins according to some rule. We will not speculate on this further, except to note that the rule in this analogy is by no means random; the rescaling of the overlap we had to perform is merely by powers of $-\log\ep$, whereas the number of causally disconnected boxes grows as $\ep^{-1}$. 

Interestingly, the Gumbel density is also the distribution of a much simpler quantity, namely the (regularized and zero mode subtracted) self-overlap $\int \d \th | \phi(\th)|^2$, as explained in \cite{AD}. 
A priori there was no reason to expect this heuristic notion of overlap to agree with the full phase-space definition (certainly we wouldn't in general), and this should presumably be taken as a further testament to the simplicity of the system we are studying.

We have also obtained an explicit expression for the triple overlap distribution (\ref{TripleDis}) which adds this system to a rather short list of models in the literature whose triple overlaps are known in closed form.

Again, we could consider first the naive  $\ep \to 0$ limit, in which this distribution approaches a delta function supported on equilateral triangles (with all three $q_a$ equal to each other), whose size tends to zero as $(-\log \ep)^{-1/2}$. While this limiting distribution is technically ultrametric, its ultrametricity is of a trivial nature, since as we have discussed above, in the same limit the regular overlap distribution also approaches a delta function, precluding any interesting phase structure. Furthermore, in a non-trivial tree the pairwise path distances between three arbitrary leaves should be allowed to take different values (only the largest two, but not necessarily all three of them should have to be equal).

However, the appropriately shifted and scaled triple overlap distribution is perfectly smooth, as shown in Figure~3, and given by a rather intriguing Kontorovich-Lebedev integral (\ref{TripleDis}). Still, it is invariant under rotations around the equilateral (i.e.~$z$) axis, which implies that isosceles triangles are not particularly preferred. If there were a non-trivial tree structure underlying these vacuum fluctuations, we wouldn't expect the triple overlap distribution to have support everywhere around the equilateral axis, but only on the three half-planes described by $q_1 = q_2 > q_3$ and permutations thereof. Thus while the rescaled triple overlap distribution certainly has a very interesting structure indicative of ergodicity breaking, we do not see evidence for a hierarchical organization of the state space\footnote{The expression (\ref{TODChar}) was subsequently rederived as the triple distance distribution of a different, simpler measure of overlap in \cite{AD}, where it was used to argue in favor of ultrametricity of the model. This was based on the observation that if one restricts to rare events in the tail of the distribution, the conditional probability $\mathcal{P}(\hat{q}_1| \hat{q}_2, \hat{q}_3)$ is peaked in the region where the three $\hat{q}$'s form isosceles triangles. However, while intriguing in its own right, this approximate localization property is clearly much weaker than ultrametricity, whose definition requires a sharp localization on a two-dimensional surface. Furthermore, it is really a property of a particular conditional distribution, rather than of the joint triple overlap distribution, since it depends crucially on the choice of conditions (which amounts to specifying a scheme for comparing isosceles triangles with others that would not be present in an ultrametric distribution). 
In this case the conditions imposed explicitly break the rotational symmetry of the joint distribution about the line of equilateral triangles, which maps isosceles triangles to non-isosceles ones. On the other hand, if we looked at a conditional distribution that respects this symmetry, e.g.~by keeping $\hat{q}_z$ (i.e.~the sum of the three sides of the triangle) constant, we would find no such approximate localization.}.

This result might be a limitation forced upon us by the low dimensionality and simple dynamics of our model. Higher dimensional de Sitter spaces may exhibit a richer structure, and we hope that the calculations presented here can be extended, at least at some level of approximation, to these cases. Other possible generalizations include introducing masses and self-interactions, or considering non-scalar fields, and we hope that the study of such systems will reveal further interesting facts about the statistical properties of inflating universes at the largest scales.

\section*{Acknowledgments}
I am very grateful to F.~Denef for introducing me to the concept of overlap distributions and giving me the idea to perform the computations presented in this note. I have enjoyed many fruitful discussions with him and had the pleasure of his collaboration during the early stages of this project. I am also indebted to D.~Anninos for sharing some of his unpublished notes on this topic with me.
Furthermore, I would like to thank I.~Klebanov for useful comments on the manuscript.
This work was partially supported by DOE grant DE-FG02-92ER40697.

\appendix

\section{Some (not so) Easy Pieces}

Here we collect some expressions that are relevant for the preliminary computations of Section \ref{Low} and might potentially lead to an alternative, simpler derivation of the main results of this paper (and generalizations thereof).

\subsection{Angular Integral for $K=3$} \label{K3Int}

Let's introduce the short-hand $\la_{ij} \equiv -\la_i \la_j / \log \ep$. We need to evaluate the angular integral in (\ref{qcubed}), which in general is not trivial.  Even though we are really interested in small values of the $\la_{ij}$, we can proceed by computing the integral for integer values of the $\la_{ij}$ first. This is simpler, since it allows a binomial expansion of the integrand, with enables us to express the answer as a multiple sum with constraints implemented by Kronecker deltas. Analytic continuation will then give us the answer for general values of the parameters.

\[ \label{3AngInt}
&&\int_0^{2\pi} \d\theta_1 \d\theta_2 \d\theta_3 \left(\sin^2 { \theta_1-\theta_2 \over 2}\right)^{\lambda_{12}}  \left(\sin^2 { \theta_1-\theta_3 \over 2}\right)^{\lambda_{13}} \left(\sin^2 { \theta_2-\theta_3 \over 2}\right)^{\lambda_{23}}  \nn \\
&=& {(2 \pi )^3 \over (-4)^{\la_{12}+\la_{13}+\la_{23}}} \sum_{k_{12}=0}^{\la_{12}} {2 \la_{12} \choose k_{12}} \sum_{k_{13}=0}^{\la_{13}} {2 \la_{13} \choose k_{13}} \sum_{k_{23}=0}^{\la_{23}} {2 \la_{23} \choose k_{23}} \nn \\
&&\times(-1)^{k_{12}+k_{13}+k_{23}}\, \de(\la_{12}-k_{12}+\la_{13}-k_{13})\,\de(-\la_{12}+k_{12}+\la_{23}-k_{23}) \nn \\
&=& {(2 \pi )^3 \over 4^{\la_{12}+\la_{13}+\la_{23}}}\sum_{k'} {(-1)^{k'} (2\la_{12})! (2\la_{13})! (2\la_{23})! \over  (\la_{12}-k')! (\la_{12}+k')!(\la_{13}-k')! (\la_{13}+k')!(\la_{23}-k')! (\la_{23}+k')!}\nn \\
&=& 8 \pi^{3/2} \,{\Ga({{1\over 2}+ \la_{12}}) \Ga({{1\over 2}+ \la_{13}}) \Ga({{1\over 2}+ \la_{23}}) \Ga(1+ \la_{12}+ \la_{13}+ \la_{23})  \over \Ga(1+ \la_{12}+ \la_{13})\Ga(1+ \la_{12}+ \la_{23})\Ga(1+ \la_{13}+ \la_{23})} \ .
\]
If we had performed the computation for general values of the $\la_{ij}$ from the start, the principal differences we would have encountered are generalized binomial sums that no longer terminate, and Kronecker deltas being replaced by sinc-functions.

\subsection{Angular Integral for $K=4$} \label{K4Int}

For the angular integral relevant to the fourth moment we would like to use the same trick as in appendix \ref{K3Int} above, namely compute for integer exponents and then analytically continue to small values of the $\la_{ij}$. In this case there are six angular factors, and four integrals, which lead to three independent Kronecker deltas. Proceeding exactly as for the $K=3$ case we find
\[
& \int_0^{2\pi} \prod_{k=1}^4 \d\theta_k \prod_{j>i \ge 1}^4 \left(\sin^2 { \theta_i-\theta_j \over 2}\right)^{\lambda_{ij}} = 
\qquad \qquad \qquad \qquad \qquad \qquad \qquad \qquad \qquad \quad \\
& {(2\pi)^4 \over 4^{\la_{12}+\la_{13}+\la_{14}+\la_{23}+\la_{24}+\la_{34}}} \sum_{k'_{12},k'_{13},k'_{23}}
 {(-1)^{{k'_{12}}+{k'_{13}}+{k'_{23}}}(2{\la_{12}})!(2{\la_{13}})!(2{\la_{14}})!(2{\la_{23}})!(2{\la_{24}})!(2{\la_{34}})! \over
({\la_{12}}-{k'_{12}})!({\la_{12}}+{k'_{12}})!({\la_{13}}-{k'_{13}})!({\la_{13}}+{k'_{13}})!({\la_{23}}-{k'_{23}})!({\la_{23}}+{k'_{23}})!} \nn \\
& \times { 1 \over
({\la_{14}}-{k'_{12}}-{k'_{13}})!({\la_{14}}+{k'_{12}}+{k'_{13}})!({\la_{24}}-{k'_{12}}+{k'_{23}})!({\la_{24}}+{k'_{12}}-{k'_{23}})!({\la_{34}}+{k'_{13}}+{k'_{23}})!({\la_{34}}-{k'_{13}}-{k'_{23}})!} \ . \qquad \quad  \nn
\]

Unfortunately, we do not know how to compute this triple sum in closed form, and without such an expression we cannot analytically continue to small values of the $\la_{ij}$. Rewriting the sums as contour integrals does also not seem to lead to an expression that readily lends itself to contour deformation and analytic continuation.

\subsection{A Modest Proposal} \label{Modest}

Of course, since we really care about small exponents we could be more modest and simply calculate the relevant integrals as a power series around $\la_{ij}=0$. If we expand in $\la_{ij}$ first we will have to compute integrals with multiple factors of 
\[ \label{FourierEdge}
\log\left(4\sin^2 {\De\th \over 2}\right) = \sum_{n=1}^\infty \left(-{2 \over n}\right) \cos n \De\th \ ,
\]
where we have indicated the Fourier series of this expression, which will be useful below.
 
Let us illustrate the procedure with the simplest case of $K=2$
\[
&&\int_0^{2\pi} \d\De\th \left(4 \sin^2 {\De\th \over 2}\right)^{\la} = \int_0^{2\pi} \d\De\th \sum_{m=0}^\infty {\la^m \over m!} \prod_{i=1}^m \sum_{n_i=1}^\infty \left(-{2 \over n_i}\right) \cos n_i \De\th \nn \\
&&= 2 \pi  \sum_{m=0}^\infty {(-\la)^m \over m!} \sum_{n_1=1}^\infty \ldots \sum_{n_m=1}^\infty \ \sum_{2^m\, \mathrm{sign\ combinations}} {\de(\pm n_1 \pm n_2 \ldots \pm n_m) \over n_1 n_2 \ldots n_m} \ .
\]
We have to sum over an $m$-dimensional integer lattice with a constraint imposed by a Kronecker delta. The coefficient at $\O(\la^m)$ should be equal to $\zeta(m)$ times a rational number.

This method generalizes straightforwardly to higher $K$: in general the number of $n_i$ factors in the denominator is equal to the total exponent of the $\la_{ij}$s and there are $K-1$ constraints, again summed over all possible sign combinations.
If we could efficiently perform such constraint $\zeta$-function type lattice sums we would have solved the problem, but in the absence of good technique to achieve this it is not clear if modesty is really advantageous in this case.

{}

\end{document}